\documentclass[a4paper,fleqn]{cas-sc}

\usepackage[numbers]{natbib}

\usepackage{lineno}
\newcommand*\patchAmsMathEnvironmentForLineno[1]{
  \expandafter\let\csname old#1\expandafter\endcsname\csname #1\endcsname
  \expandafter\let\csname oldend#1\expandafter\endcsname\csname end#1\endcsname
  \renewenvironment{#1}
     {\linenomath\csname old#1\endcsname}
     {\csname oldend#1\endcsname\endlinenomath}}
\newcommand*\patchBothAmsMathEnvironmentsForLineno[1]{
  \patchAmsMathEnvironmentForLineno{#1}
  \patchAmsMathEnvironmentForLineno{#1*}}
\AtBeginDocument{
\patchBothAmsMathEnvironmentsForLineno{equation}
\patchBothAmsMathEnvironmentsForLineno{align}
\patchBothAmsMathEnvironmentsForLineno{flalign}
\patchBothAmsMathEnvironmentsForLineno{alignat}
\patchBothAmsMathEnvironmentsForLineno{gather}
\patchBothAmsMathEnvironmentsForLineno{multline}
}

\def\tsc#1{\csdef{#1}{\textsc{\lowercase{#1}}\xspace}}
\tsc{WGM}
\tsc{QE}

\begin{document}
\let\WriteBookmarks\relax
\def\floatpagepagefraction{1}
\def\textpagefraction{.001}

\shorttitle{Novel phase-retrieval algorithm from a single-shot image for X-ray schlieren microscopy}

\shortauthors{R. Nishimura, et al.}  

\title [mode = title]{Development of a new phase-retrieval algorithm from a single-shot image for X-ray schlieren microscopy}

\tnotemark[1] 

\tnotetext[1]{} 

%

\author[1,2]{Ryutaro NISHIMURA}[orcid=0000-0002-3156-7347]
\cormark[1]
\ead{ryutaro.nishimura@kek.jp}
\credit{Conceptualization of this study, Data curation, Formal analysis, Investigation, Methodology, Software, Visualization, Writing - original draft preparation, Writing - review and editing}

\author[1]{Yoshio SUZUKI}[]
\credit{Conceptualization of this study, Data curation, Formal analysis, Investigation, Methodology, Project administration, Supervision, Validation, Visualization, Writing - original draft preparation, Writing - review and editing}

\author[1,2]{Hiroshi SUGIYAMA}[]
\credit{Funding acquisition, Project administration, Supervision, Validation, Writing - review and editing}

\author[1,2]{Daisuke WAKABAYASHI}[]
\credit{Conceptualization of this study, Data curation, Funding acquisition, Investigation, Resources, Supervision, Validation, Writing - review and editing}

\author[1,2]{Yuki SHIBAZAKI}[]
\credit{Data curation, Funding acquisition, Investigation, Resources, Supervision, Validation, Writing - review and editing}

\author[1,2,3]{Keiichi HIRANO}[]
\credit{Conceptualization of this study, Data curation, Funding acquisition, Project administration, Resources, Supervision, Validation, Writing - review and editing}

\author[1,2]{Noriyuki IGARASHI}[]
\credit{Data curation, Funding acquisition, Investigation, Project administration, Resources, Supervision, Validation, Writing - review and editing}

\author[1,2,4]{Nobumasa FUNAMORI}[]
\credit{Funding acquisition, Project administration, Supervision, Writing - review and editing}

\affiliation[1]{organization={Institute of Materials Structure Science, High Energy Accelerator Research Organization (KEK)},
            addressline={}, 
            city={Tsukuba},
            postcode={305-0801}, 
            state={Ibaraki},
            country={Japan}}
\affiliation[2]{organization={Materials Structure Science Program, Graduate Institute for Advanced Studies, Graduate University for Advanced Studies (SOKENDAI)},
            addressline={}, 
            city={Tsukuba},
            postcode={305-0801}, 
            state={Ibaraki},
            country={Japan}}
\affiliation[3]{organization={Graduate School of Pure and Applied Sciences},
            addressline={}, 
            city={Tsukuba},
            postcode={305-8571}, 
            state={Ibaraki},
            country={Japan}}
\affiliation[4]{organization={Department of Earth and Planetary Science, University of Tokyo},
            addressline={}, 
            city={Bunkyo},
            postcode={113-0033}, 
            state={Tokyo},
            country={Japan}}

\cortext[1]{Corresponding author}


\begin{abstract}
In this paper, a new phase-retrieval algorithm from an X-ray schlieren image is proposed. 
The schlieren method allows phase-contrast imaging with an objective lens and a knife-edge filter placed at the back focal plane of the objective.
This method finds a wide range of applications in the visible-light region for transparent specimen visualization.
The schlieren contrast does not directly correspond to the phase shift. 
However, the phase map can be reconstructed from a single-shot schlieren image of a transparent and weak-phase object using the filtered Fourier transform method. 
A proof-of-principle experiment was performed in the hard-X-ray region at the AR-NE1A beamline of the Photon Factory facility at the High Energy Accelerator Research Organization (KEK). 
\end{abstract}

\begin{highlights}
\item A new phase-retrieval algorithm from an X-ray schlieren image is proposed. 
\item In the algorithm, the phase map can be reconstructed from a single-shot schlieren image.
\item A transparent and weak-phase object is imaged via the filtered Fourier transform method. 
\item A PoP experiment was performed in the hard-X-ray region at the AR-NE1A beamline of KEK. 
\end{highlights}


\maketitle

\section{Introduction}

Generally, the phase contrast generated by the real part of the refractive index in a hard-X-ray region is significantly higher than the absorption contrast generated by its imaginary part \cite{Momose1}. 
In the X-ray region, numerous phase-contrast methods have been developed and applied to microscopy, such as Zernike phase-contrast microscopy \cite{Schmahl1, Takeuchi1}, the Bonse-Hart crystal optics interferometer \cite{Momose2}, the propagation-based phase-contrast method (defocusing contrast in full-field imaging microscopy) \cite{Snigirev1, Wilkins1}, and the grating interferometer method (Talbot interferometer) \cite{Momose3}. 
Additionally, the possibility of phase map reconstruction from propagation-based phase-contrast images has been demonstrated \cite{Paganin1, Suzuki1}. 
Several types of holographic imaging that are essentially phase-sensitive imaging have also been developed in the soft- and hard-X-ray regions, including inline holography \cite{Aoki1, Jacobsen1}, Fourier transform holography with Fresnel zone plate optics \cite{MacNulty1, Leitenberger1}, two-beam interferometers with refractive prisms \cite{Suzuki2, Suzuki3}, and total reflection mirrors \cite{Suzuki4, Suzuki5}. 

The propagation-based method (refraction-enhanced contrast) is the most widely used technique for phase-contrast imaging in the hard-X-ray region. 
An advantage of the refraction-enhanced approach is that no additional optical devices are required, and neither high spatial coherency nor high temporal coherency are necessary.

\begin{figure}
\centering
\includegraphics[width=\linewidth]{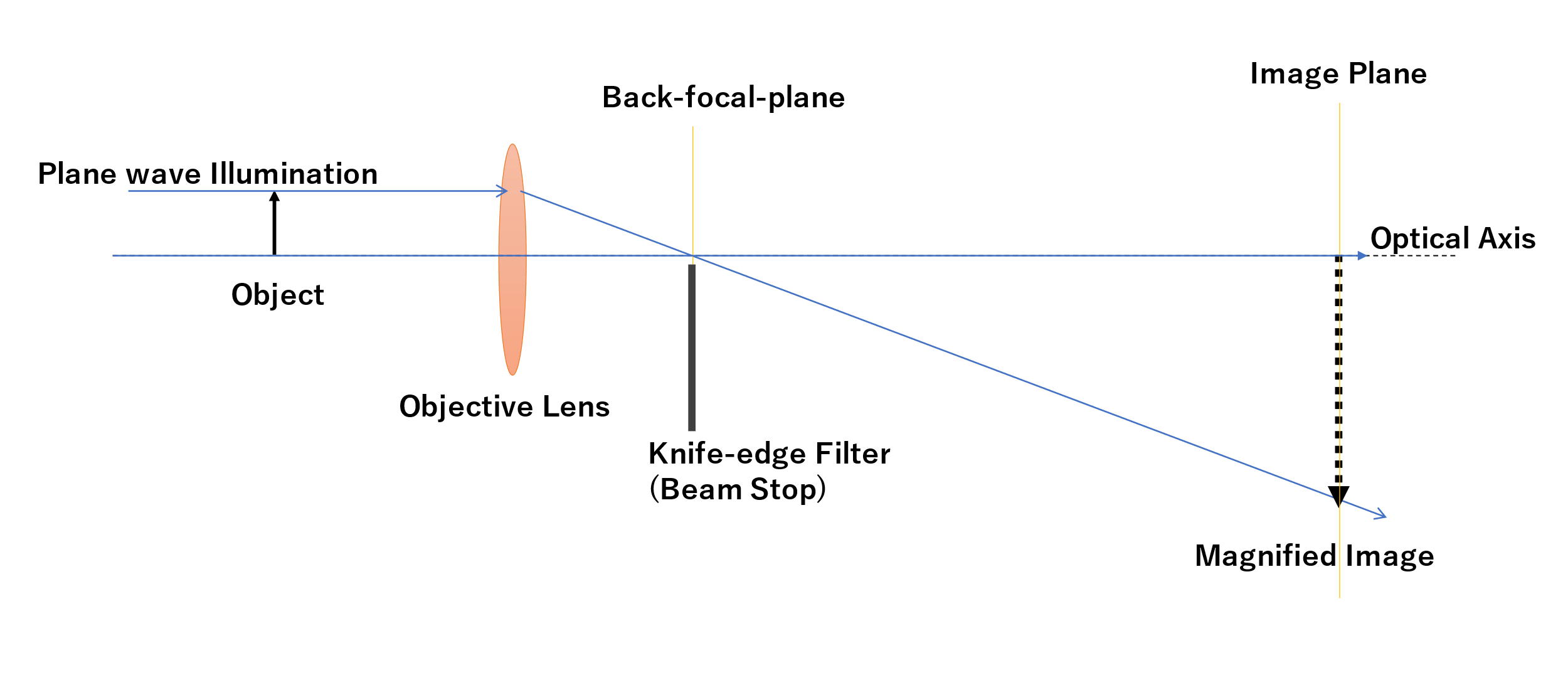}
\caption{\label{fig1} Optical system for schlieren microscopy.}
\end{figure}

In the visible-light region, the simple and conventional schlieren imaging method \cite{Born1} is widely used to visualize transparent objects with a non-uniform optical thickness.
The optical system of the traditional schlieren method is schematically shown in Fig. \ref{fig1}. 
When an object is illuminated by a parallel beam, the beam penetrates the object directly and creates a single spot on the back focal plane of an objective lens.
The diffracted waves generate several spots around the focal point, and their distance from the central spot is proportional to the spatial frequency of the corresponding structure of the object. 
The plane of focus is referred to as the Fourier plane.
The schlieren contrast is generated by excluding one side of the diffracted waves by placing a single-blade beam stop (knife-edge filter) at the back focal plane of the objective lens.
Generally, the contrast of an image appears as an edge-enhanced bright-dark structure known as the schlieren texture. 
Therefore, the schlieren method is not quantitative, but an easy and robust technique as the phase-contrast image can be obtained by simply placing a knife-edge filter at the back focal plane. Furthermore, the spatial resolution of the schlieren image is essentially the same as that of a conventional absorption-contrast image. 

Recently, Watanabe and Aoki \cite{Watanabe1} reported a differential phase-contrast method based on the linear motion of the knife-edge filter placed at the back focal plane (the Foucault knife-edge scanning filter method). 
A phase map was obtained from the line integral of the measured differential phase images.
However, capturing an image during continuous motion or a series of images during step motion of the knife-edge filter poses several challenges. 
For example, its application to tomography is difficult owing to the long measurement time taken for tomography scans due to complicated scanning procedures.
In this paper, we propose a new phase-retrieval method from a single schlieren microscopy image in the hard-X-ray region. 
The proposed imaging method is simple and fast because the phase map can be reconstructed from a single image captured without additional scanning. 
Therefore, the proposed method is expected to find applications in high-speed live imaging and tomography. 
Although our method is suitable only for transparent (negligible absorption) and weak-phase objects, the conditions can be generally satisfied for light element specimens with thin thicknesses in the hard-X-ray region. 
Additionally, feasibility studies were conducted with a Fresnel zone plate (FZP) as the objective lens in the hard-X-ray region (X-ray energy of 10 keV) at the AR-NE1A beamline of the Photon Factory at the High Energy Accelerator Research Organization (KEK).

\section{Schlieren imaging and phase retrieval theory}

\subsection{Basic schlieren imaging theory}

To consider the simple case of a phase-contrast object, a transparent object with a weak phase contrast is assumed. 
The transmission function is expressed as in Eq. \ref{eq1}: 
\begin{equation}
\label{eq1}
E = \exp^{i\phi(x)}
\end{equation}
where $x$ denotes a position of the phase-contrast direction on the input field and $\phi(x)$ denotes a real function that expresses the phase-shift map. 
The input field is assumed to be unity and the magnitude of $\phi$ is assumed to be sufficiently small compared to unity. 
Then, ignoring the higher-order terms, the transmission function can be written as follows: 
\begin{equation}
\label{eq2}
E \simeq 1 + i \phi(x)
\end{equation}
The Fourier transform of the phase-shift map $\phi(x)$ is defined as follows:
\begin{equation}
\label{eq3}
\Phi(k) = \int_{-\infty}^{\infty} i \phi (x) \exp^{-2 \pi ikx} dx
\end{equation}
where $k$ denotes the wavenumber of the reciprocal space. 
Therefore, the transmission function can be described using an inverse Fourier transform, as follows:
\begin{equation}
\label{eq4}
E \simeq 1 + i \phi(x) = 1 + \int_{-\infty}^{\infty} \Phi (k) \exp^{2 \pi ikx} dk
\end{equation}
The phase shift $\phi$ is defined as a purely real value. 
Therefore, the imaginary part of $\phi$ must be zero, such that
\begin{equation}
\label{eq5}
\phi(x) - \phi^* (x) = 0
\end{equation}
Here, $\phi^*$ denotes the complex conjugate of $\phi$. 
Using the inverse Fourier transform of $\Phi(k)$, Eq. \ref{eq6} and Eq. \ref{eq7} can be derived. 
\begin{equation}
\label{eq6}
\frac{1}{i} \int_{-\infty}^{\infty} \Phi(k) \exp^{2 \pi ikx} dk - \frac{1}{-i} \int_{-\infty}^{\infty} \Phi^* (k) \exp^{-2 \pi ikx} dk = 0
\end{equation}
\begin{equation}
\label{eq7}
\therefore \int_{-\infty}^{\infty} \Phi(k) \exp^{2 \pi ikx} dk + \int_{-\infty}^{\infty} \Phi^* (-k) \exp^{2 \pi ikx} dk = 0
\end{equation}
Taking the Fourier transform of Eq. \ref{eq7}, we have the following:
\begin{equation}
\label{eq8}
\int_{-\infty}^{\infty} \exp^{-2 \pi ikx} dx \int_{-\infty}^{\infty} \Phi(k) \exp^{2 \pi ikx} dk + \int_{-\infty}^{\infty} \exp^{-2 \pi ikx} dx \int_{-\infty}^{\infty} \Phi^* (-k) \exp^{2 \pi ikx} dk = 0
\end{equation}
Here, $\Phi(k)$ must satisfy the following condition \cite{Born1}:
\begin{equation}
\label{eq9}
\Phi(k) + \Phi^* (-k) = 0
\end{equation}
The schlieren contrast can be generated by excluding one side of the diffracted waves \cite{Born1}, as follows:
\begin{equation}
\label{eq10}
E = 1 + \int_{0}^{\infty} \Phi(k) \exp^{2 \pi ikx} dk
\end{equation}
The intensity observed at the detector can be expressed by the square of the absolute value of the field $E$ as in Eq. \ref{eq11}.
\begin{eqnarray}
\label{eq11}
I & = & |E|^2 = E \times E^* \nonumber \\ 
 & = & 1 + \left[ \int_{0}^{\infty} \Phi(k) \exp^{2 \pi ikx} dk \right] + \left[ \int_{0}^{\infty} \Phi(k) \exp^{2 \pi ikx} dk \right]^* \nonumber \\ 
& & + \left[ \int_{0}^{\infty} \Phi(k) \exp^{2 \pi ikx} dk \right] \times \left[ \int_{0}^{\infty} \Phi(k) \exp^{2 \pi ikx} dk \right]^*
\end{eqnarray}
By ignoring the higher-order terms, the image intensity $I$ can be approximated as follows:
\begin{eqnarray}
\label{eq12}
I & \simeq & 1 + \left[ \int_{0}^{\infty} \Phi(k) \exp^{2 \pi ikx} dk \right] + \left[ \int_{0}^{\infty} \Phi(k) \exp^{2 \pi ikx} dk \right]^* \nonumber \\ 
& = & 1 + \left[ \int_{0}^{\infty} \Phi(k) \exp^{2 \pi ikx} dk \right] + \left[ \int_{0}^{\infty} \Phi^* (k) \exp^{-2 \pi ikx} dk \right]
\end{eqnarray}
Subsequently, using Eq. \ref{eq9}, we obtain Eq. \ref{eq13}.
\begin{eqnarray}
\label{eq13}
I & \simeq & 1 + \left[ \int_{0}^{\infty} \Phi(k) \exp^{2 \pi ikx} dk \right] + \left[ \int_{0}^{\infty} -\Phi(-k) \exp^{-2 \pi ikx} dk \right] \nonumber \\ 
& = & 1 + \left[ \int_{0}^{\infty} \Phi(k) \exp^{2 \pi ikx} dk \right] + \left[ \int_{-\infty}^{0} -\Phi(k) \exp^{2 \pi ikx} dk \right]
\end{eqnarray}
Finally, the contrast function of the schlieren image, $\Delta I(x)$, can be written as
\begin{equation}
\label{eq14}
\Delta I(x) = I(x) - 1 \simeq \int_{-\infty}^{\infty} sign(k) \Phi(k) \exp^{2 \pi ikx} dk
\end{equation}
where $sign(k)$ is defined as follows:
\begin{equation}
\label{eq15}
sign(k) = \left\{
 \begin{array}{l}
  +1 \;\;\; (k>0)\\
  -1 \;\;\; (k<0)
 \end{array}
\right.
\end{equation}
Hence, schlieren-contrast intensity modulation can be considered as a type of filtered image of the phase-shift map with a frequency filter of $+1$ for $k > 0$ and $-1$ for $k < 0$. 
This frequency filter is similar to the filter used in the Fourier representation of the Hilbert transform \cite{Hilbert1}. 
The spatial resolution limit of the reconstructed phase image is essentially the same as that of the schlieren- and absorption-contrast images.

\subsection{Phase retrieval from a schlieren image}

By taking the Fourier transform of Eq. \ref{eq14} on both the sides, we obtain Eq. \ref{eq16}, as follows:
\begin{equation}
\label{eq16}
\int_{-\infty}^{\infty} \Delta I(x) \exp^{-2 \pi ikx} dx = \int_{-\infty}^{\infty} \exp^{-2 \pi ikx} dx \int_{-\infty}^{\infty} sign(k) \Phi(k) \exp^{2 \pi ikx} = sign(k) \Phi(k)
\end{equation}
such that we obtain Eq. \ref{eq17}. 
\begin{equation}
\label{eq17}
\Phi(k) = sign(k) \int_{-\infty}^{\infty} \Delta I(x) \exp^{-2 \pi ikx} dx
\end{equation}
Using the definition of $\Phi(k)$, Eq. \ref{eq17} can be rewritten as
\begin{equation}
\label{eq18}
\Phi(k) = \int_{-\infty}^{\infty} i \phi(x) \exp^{-2 \pi ikx} dx = sign(k) \int_{-\infty}^{\infty} \Delta I(x) \exp^{-2 \pi ikx} dx
\end{equation}
Taking the inverse Fourier transform of Eq. \ref{eq18}, the phase function $i \phi(x)$ can be written as follows:
\begin{eqnarray}
\label{eq19}
i \phi(x) & = & \int_{-\infty}^{\infty} \exp^{2 \pi ikx} dk \int_{-\infty}^{\infty} i \phi(x) \exp^{-2 \pi ikx} dx \nonumber \\ 
& = & \int_{-\infty}^{\infty} \exp^{2 \pi ikx} dk \: sign(k) \int_{-\infty}^{\infty} \Delta I(x) \exp^{-2 \pi ikx} dx
\end{eqnarray}
Therefore, the phase-shift map can be obtained by the Fourier transform of the intensity modulation, followed by the inverse Fourier transform with a simple filter function of $+1$ for $k > 0$ and $-1$ for $k < 0$. 

\subsection{Expansion to the partial coherent condition}

The theory of schlieren imaging and phase retrieval under parallel-beam illumination conditions are discussed in this section. 
Usually, the actual illumination is partially coherent. The expansion of phase retrieval to the partially coherent illumination condition has been demonstrated.
To consider a partially coherent state, an illuminating beam consisting of an ensemble of independent beams that each have a different inclination with respect to the optical axis is considered. 
Thereafter, the signal intensity is obtained by the summation of the intensities of each plane-wave beam. 
When the incident wave is parallel and the inclination angle to the optical axis is $\theta$, the schlieren contrast can be described as follows:
\begin{equation}
\label{eq20}
I(x) = 1 + \left[ \int_{k_0}^{\infty} \Phi(k) \exp^{2 \pi ikx} dk \right] + \left[ \int_{-\infty}^{-k_0} -\Phi(k) \exp^{2 \pi ikx} dk \right]
\end{equation}
Here, $k_0$ is defined as follows:
\begin{equation}
\label{eq21}
\frac{\sin \theta}{\lambda} = k_0
\end{equation}
where $\lambda$ is the X-ray wavelength and ${\lambda}/{\sin \theta}$ represents the period of the object structure.
Therefore, an object structure with a spatial frequency lesser than $k_0$ does not yield contrast in the schlieren image. 
Conversely, if the object does not have a structure with a spatial frequency lesser than $k_0$, the schlieren contrast is essentially the same as that in the normal incidence condition. 

\begin{figure}
\centering
\includegraphics[width=\linewidth]{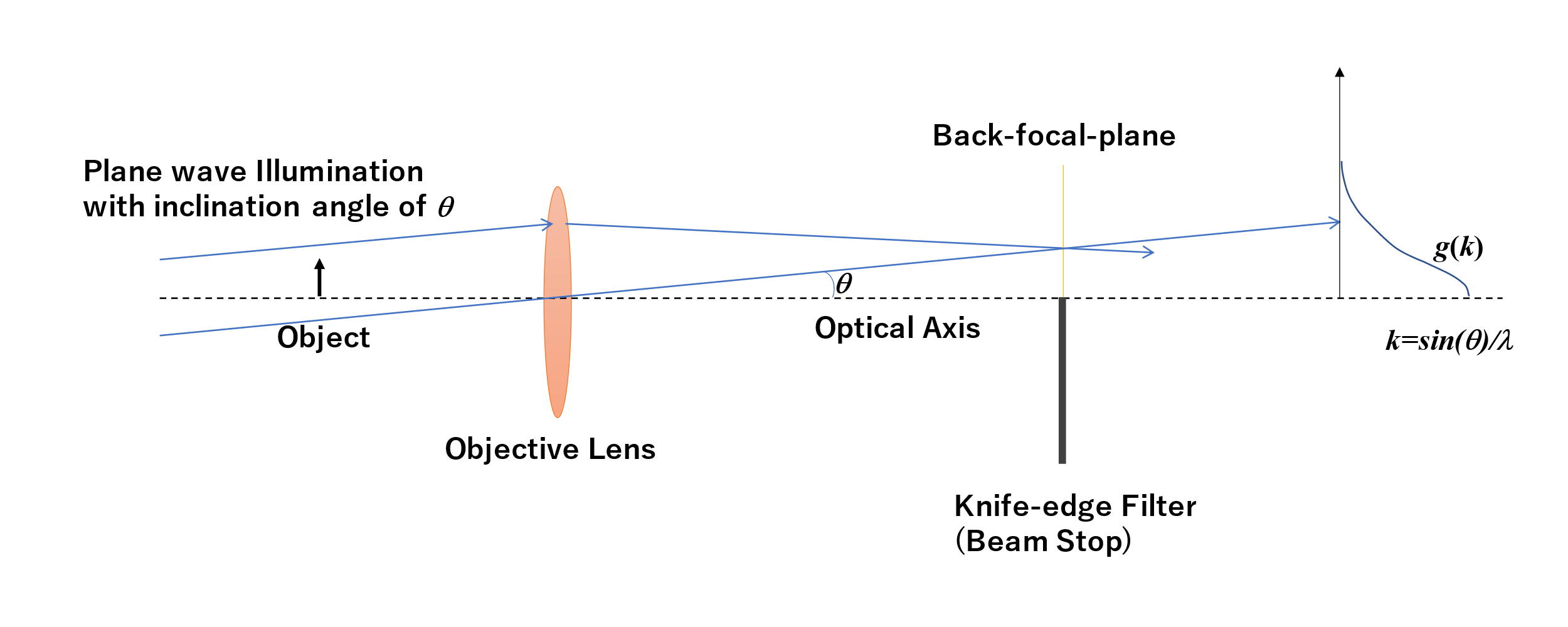}
\caption{\label{fig2} Schlieren method with inclined plane wave illumination.}
\end{figure}

When a part of the illuminating beam arrives at the object/objective, the angular distribution of the incident beam is assumed to be $g(k_0)$, as shown in Fig. \ref{fig2}, and the beam stop is placed at $\theta = 0$. The intensity modulation can be described as follows:
\begin{equation}
\label{eq22}
\Delta I(x) = I(x) - 1 \simeq \int_{-\infty}^{\infty} G(k) \Phi(k) \exp^{2 \pi ikx} dk
\end{equation}
where $G(k)$ is defined by the integral of $g(k)$, as in Eqs. \ref{eq23} and \ref{eq24}.
\begin{eqnarray}
\label{eq23}
G(k) & \equiv & \int_{0}^{k} g(k') dk' \;\;\; (k>0) \\
\label{eq24}
G(k) & \equiv & -G(|k|) \;\;\; (k<0)
\end{eqnarray}
Thereafter, $G(k)$ is assumed to be normalized as in Eq. \ref{eq25}. 
\begin{equation}
\label{eq25}
G(\infty) \equiv \int_{0}^{\infty} g(k) dk = 1
\end{equation}
Then, the phase map can be reconstructed using the integral of the angular distribution function $G(k)$, as follows:
\begin{eqnarray}
\label{eq26}
i \phi(x) & = & \int_{-\infty}^{\infty} \exp^{2 \pi ikx} dk \int_{-\infty}^{\infty} i \phi(x) \exp^{-2 \pi ikx} dx \nonumber \\
 & = & \int_{-\infty}^{\infty} \exp^{2 \pi ikx} dk \frac{1}{G(k)} \int_{-\infty}^{\infty} \Delta I(x) \exp^{-2 \pi ikx} dx
\end{eqnarray}
Hence, if the angular distribution function is exactly known, the phase map can be retrieved from the measured single image using the schlieren method. \\
Apparently, when $g(k)$ is a delta function, the intensity modulation can be simply written as
\begin{equation}
\label{eq27}
\Delta I(x) \simeq \int_{-\infty}^{\infty} sign(k) \Phi(k) \exp^{2 \pi ikx} dk
\end{equation}
Eq. \ref{eq27} represents a coherent illumination. 
Conversely, for a sufficiently low-coherent illumination, where the angular distribution of the illuminating beam is uniform such that $g(k)$ is constant for $k < k_{max}$, where $k_{max}$ is the maximum value of $k$, the integral $G(k)$ can be written as in Eq. \ref{eq28}. 
\begin{equation}
\label{eq28}
G(k) = \frac{k}{k_{max}}
\end{equation}
Then, the schlieren contrast can be expressed as follows:
\begin{eqnarray}
\label{eq29}
\Delta I(x) & \simeq & \int_{-\infty}^{\infty} \frac{k}{k_{max}} \Phi(k) \exp^{2 \pi ikx} dk \nonumber \\
 & = & \frac{d}{dx} \int_{-\infty}^{\infty} \frac{1}{2 \pi i k_{max}} \Phi(k) \exp^{2 \pi ikx} dk \nonumber \\
 & = & \frac{1}{2 \pi k_{max}} \frac{d}{dx} \phi(x)
\end{eqnarray}
Therefore, for a low-coherent illumination condition, the schlieren contrast is proportional to the differential of the wavefront. Additionally, an exact phase retrieval is apparently possible using a simple integral of the measured schlieren image. 

\subsection{Correction for a finite imaging plane}

In the proposed phase-retrieval theory, the field of the imaging plane is assumed to be infinite. 
However, in an actual experimental setup, the imaging plane is limited by optics and detectors. 
In this situation, in the phase retrieval process, the edge region of the limited imaging plane behaves as a large dummy contrast which has the amplitude from zero to some finite value (usually non-phase shift background value). 
Then, without correction, it appears as a striped pattern or an undulating pattern artifact in the phase-retrieved image. 
Thus, a padding region filled with a constant value equivalent to a non-phase shift is added to each edge of the input image (intensity-normalized). 
This padding region is to ensure that the dummy contrast does not affect the sample image that phase-retrieved from the original imaging plane in the phase-retrieved image after phase recovery. 
The array size of the total image, including the padding region, should be at least eight times larger than the size of the phase-retrieval direction of the input image.

\section{Experimental setup}

\begin{figure}
\centering
\includegraphics[width=\linewidth]{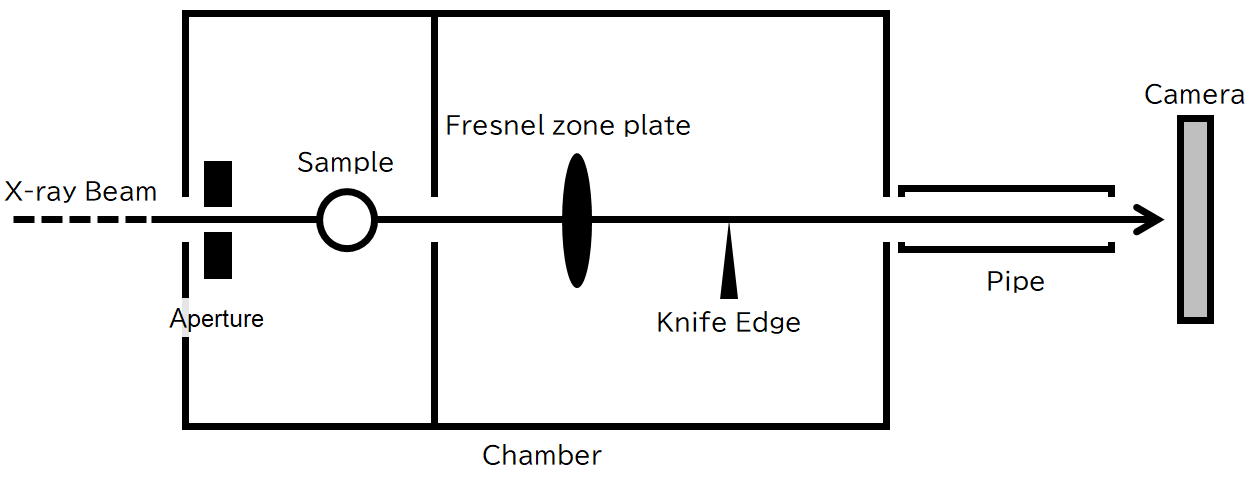}
\caption{\label{fig3} Schematic of the experimental setup for the X-ray microscope installed at the AR-NE1A beamline. The sample-to-detector distance is approximately 6.9 m, and the vacuum duct between the XRM chamber and detector is approximately 5 m.}
\end{figure}

The experimental setup of the X-ray microscope was installed at the AR-NE1A beamline of the Photon Factory facility at KEK. 
The installed experimental setup (Fig. \ref{fig3}) was essentially the same as that used for our previous experiment \cite{Wakabayashi1}, and is briefly explained next.
 
The X-ray beam from a multi-pole wiggler is monochromatized by passing through the Si 111 double-crystal monochromator, and collimated by the variable-bent double multi-layer mirror optics configured in a crossed-mirror geometry (Kirkpatrick-Baez optics) to achieve a nearly parallel beam illumination. 
Then, the X-ray beam is collimated by the aperture and illuminates the sample. 
Only the lower half of the FZP is irradiated for diffraction-order-sorting in the parallel beam illumination.
The X-ray beam that passes through the sample is magnified by the FZP, and partly cut-off by the knife edge placed at the back-focal-plane of the objective FZP to obtain the schlieren phase contrast. 
Finally, the X-ray beam passes through the vacuum pipe (approximately 5-m long) and is imaged by the X-ray camera. 
The distance between the object and X-ray camera is set to approximately 6.9 m. 
A complementary metal-oxide-semiconductor camera (C12849-111U, Hamamatsu Photonics) with a pixel format of 2048 $\times$ 2048, a pixel size of 6.5 {\textmu}m, a 10-{\textmu}m-thick gadolinium oxysulfide (P43) scintillator, and 1:1 fiber optics, was used as the X-ray camera. 
In the experiment, a single FZP optics was employed. 
The FZP used in the current experiment was fabricated by Applied Nanotools (Canada), and has a diameter of 550 {\textmu}m and outermost zone width of 50 nm. 
The zone material used was 0.5-{\textmu}m-thick gold. 
In the experiment, a 10-keV monochromatic X-ray beam was used. 
Therefore, the focal length of the FZP is calculated to be 222 mm, and the magnification is calculated to be $\times$29. 
The actual magnification was calibrated by measuring the calibration standard (CAL21HEI2, Applied Nanotools), and the magnification of the X-ray optics is measured to be $\times$28.5, which is consistent with the calculated value. 
The demonstration sample used in the proof-of-principle experiment, siemens-star and line-and-space patterns, was also calibration standard made of 600-nm-thick gold. 

\section{Results and Discussion}

\begin{figure}
\centering
\includegraphics[width=\linewidth]{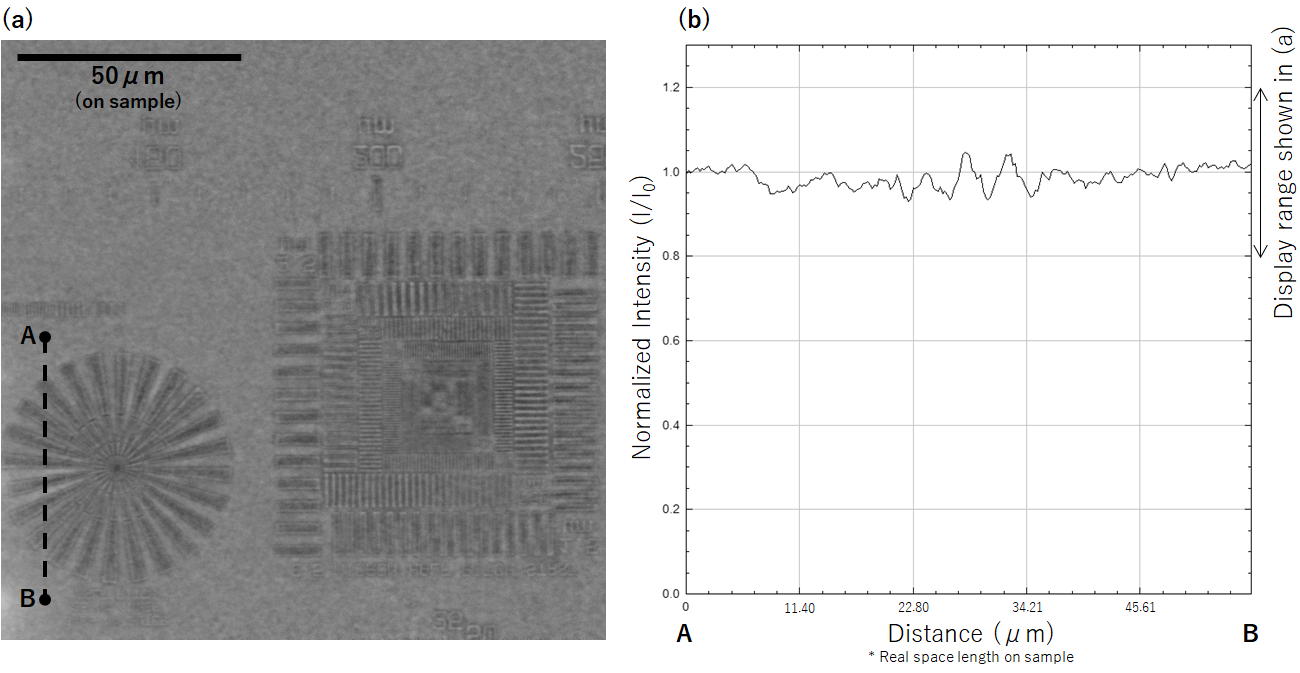}
\caption{\label{fig_abs_data} (a) Intensity-normalized absorption image of the siemens-star and line-and-space patterns following the calibration standard. The total exposure time is 50 s (5 s $\times$ 10 images) for a 15-mA stored current. (b) Intensity profile at line A-B.}
\end{figure}
\begin{figure}
\centering
\includegraphics[width=\linewidth]{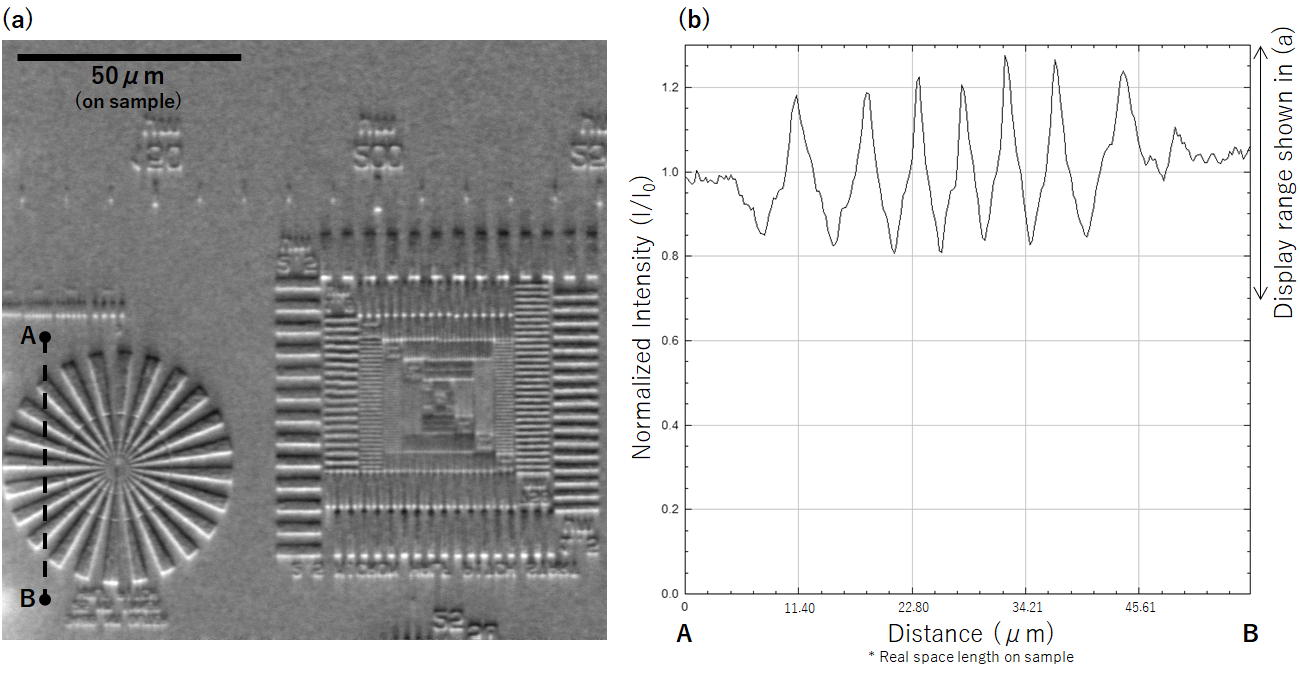}
\caption{\label{fig_sch_abs_data} (a) Intensity-normalized schlieren phase-contrast image of the siemens-star and line-and-space patterns following the calibration standard. The total exposure time is 50 s (5 s $\times$ 10 images) for a 15-mA stored current. (b) Intensity profile at line A-B.}
\end{figure}
\begin{figure}
\centering
\includegraphics[width=\linewidth]{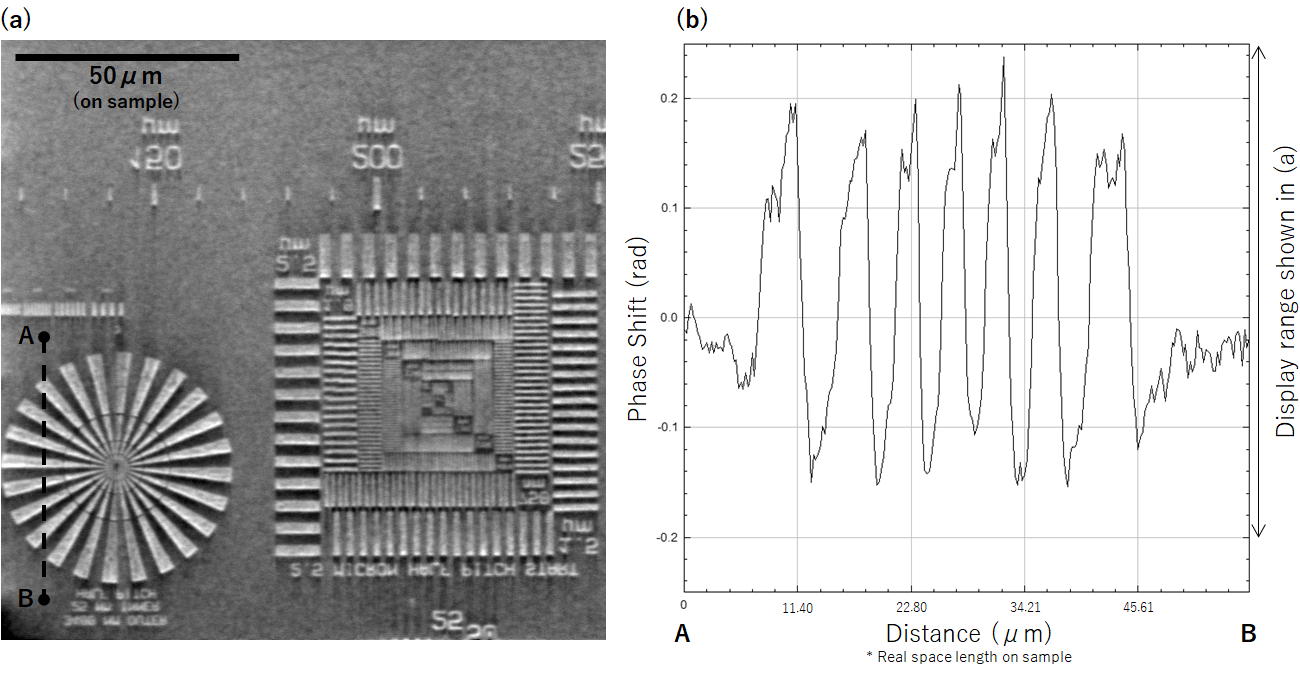}
\caption{\label{fig_sch_phase_ret_data} (a) Phase-retrieval image of the siemens-star and line-and-space patterns following the calibration standard. (b) Phase-shift profile at line A-B.}
\end{figure}

Figs. \ref{fig_abs_data}--\ref{fig_sch_phase_ret_data} present the experimental results of the demonstration. 
The total exposure time of each images was 50 s (5 s $\times$ 10 images) for a 15-mA stored current. 
Fig. \ref{fig_abs_data} (a) shows the absorption image (intensity-normalized) of the siemens-star and line-and-space patterns following the calibration standard, while Fig. \ref{fig_abs_data} (b) presents the intensity profile at line A-B in Fig. \ref{fig_abs_data} (a). 
The absorption image clearly showed that the calibration standard could be regarded as a practical pure-phase object for the given condition. 
Fig. \ref{fig_sch_abs_data} (a) shows the schlieren phase-contrast image (intensity-normalized) of the calibration standard, which was obtained by adding a simple knife-edge filter to the back focal plane; the contrast direction was vertical to the figures. 
Fig. \ref{fig_sch_abs_data} (b) presents the intensity profile at line A-B in Fig. \ref{fig_sch_abs_data} (a). 
Fig. \ref{fig_sch_phase_ret_data} (a) illustrates the phase-retrieval image processed from Fig. \ref{fig_sch_abs_data} (a) using our proposed phase-retrieval algorithm, while Fig. \ref{fig_sch_phase_ret_data} (b) presents the phase-shift profile at line A-B in Fig. \ref{fig_sch_phase_ret_data} (a). 
The phase-shift map due to the sample structures was confirmed to be reconstructed, as shown in Fig. \ref{fig_sch_phase_ret_data}, while Fig. \ref{fig_sch_abs_data} shows the typical schlieren texture with the bright and dark pair contrast at the edge of the object structure.
 
The reconstructed phase-retrieval image had a reasonable profile; a positive phase shift occurred in the thin metal region, while other regions showed a negative or no value. 
Some negative-value regions were possibly caused by optic incoherence. 
The phase shift was approximately 0.3 rad (from the negative-value region to the thin-metal region). 
Contrarily, the estimated phase shift was 0.909 rad (calculated from the thickness and material of the metal region: 600 nm of gold). 
The discrepancy could be partially attributable to the structures of the test samples. 
Although the absorption was negligible, as shown in Fig. \ref{fig_abs_data}, the weak phase-contrast condition was not fully satisfied. 
A phase shift lower than the calculated value, with a known optical constant, could also have been caused by insufficient coherence of the illuminating beam or stray light of the image-forming optics or both.
Further exact phase retrieval was possible by adjusting the filter function and improving the optics.
However, these measurements confirmed that our phase-retrieval method was effective and applicable to phase imaging in hard-X-ray full-field microscopy. 

\section{Conclusions}

In this paper, we propose a new phase-retrieval algorithm for a weak-phase object from a single-shot X-ray schlieren image. 
A proof-of-principle experiment was performed in the hard-X-ray region at the AR-NE1A beamline of the Photon Factory facility at KEK. The conventional schlieren method was found to be applicable for phase-contrast imaging in the hard-X-ray region, and the phase map was retrieved from the measured single-shot schlieren image using the proposed phase-retrieval algorithm. 
The schlieren method was applicable to optical systems with FZP objectives and optical devices such as compound refractive lenses. 
The single-shot phase-contrast image required by the proposed algorithm could be obtained by including a simple knife-edge filter to the back focal plane without step or continuous motion of the filter. 
Therefore, the proposed phase-retrieval algorithm is expected to find use in high-resolution phase-contrast imaging and phase-contrast microtomography. 

\section{Acknowledgements}

This study was conducted at the AR-NE1A beamline of the Photon Factory facility at KEK (Proposal No. 2021PF-S001). 




\section{}\label{}

\printcredits

\bibliographystyle{model1-num-names}

\bibliography{opt_manuscript_reference.bib}

\begin{thebibliography}{21}
\expandafter\ifx\csname natexlab\endcsname\relax\def\natexlab#1{#1}\fi
\providecommand{\bibinfo}[2]{#2}
\ifx\xfnm\relax \def\xfnm[#1]{\unskip,\space#1}\fi
\bibitem[{Momose(2005)}]{Momose1}
\bibinfo{author}{A.~Momose},
\newblock \bibinfo{title}{Recent advances in x-ray phase imaging},
\newblock \bibinfo{journal}{Jpn. J. Appl. Phys.} \bibinfo{volume}{44}
  (\bibinfo{year}{2005}) \bibinfo{pages}{6355--6367}.
\bibitem[{Schmahl. et~al.(1994)Schmahl., Rudolph, Schneider, Guttmann, and
  Niemann}]{Schmahl1}
\bibinfo{author}{G.~Schmahl.}, \bibinfo{author}{D.~Rudolph},
  \bibinfo{author}{G.~Schneider}, \bibinfo{author}{P.~Guttmann},
  \bibinfo{author}{B.~Niemann},
\newblock \bibinfo{title}{Phase contrast x-ray microscopy studies},
\newblock \bibinfo{journal}{Optik} \bibinfo{volume}{97} (\bibinfo{year}{1994})
  \bibinfo{pages}{181--182}.
\bibitem[{Takeuchi et~al.(2009)Takeuchi, Uesugi, and Suzuki}]{Takeuchi1}
\bibinfo{author}{A.~Takeuchi}, \bibinfo{author}{K.~Uesugi},
  \bibinfo{author}{Y.~Suzuki},
\newblock \bibinfo{title}{Zernike phase-contrast x-ray microscope with
  pseudo-kohler illumination generated by sectored (polygon) condenser plate},
\newblock \bibinfo{journal}{J. Phys.: Conf. Ser.} \bibinfo{volume}{186}
  (\bibinfo{year}{2009}) \bibinfo{pages}{012020}.
\bibitem[{Momose(1995)}]{Momose2}
\bibinfo{author}{A.~Momose},
\newblock \bibinfo{title}{Demonstration of phase-contrast x-ray computed
  tomography using an x-ray interferometer},
\newblock \bibinfo{journal}{Nucl. Instrum. Meth. A} \bibinfo{volume}{352}
  (\bibinfo{year}{1995}) \bibinfo{pages}{622--628}.
\bibitem[{Snigirev et~al.(1995)Snigirev, Snigireva, Kohn, Kuznetsw, and
  Schelokov}]{Snigirev1}
\bibinfo{author}{A.~Snigirev}, \bibinfo{author}{I.~Snigireva},
  \bibinfo{author}{V.~Kohn}, \bibinfo{author}{S.~Kuznetsw},
  \bibinfo{author}{I.~Schelokov},
\newblock \bibinfo{title}{On the possibilities of x-ray phase contrast
  microimaging by coherent high-energy synchrotron radiation},
\newblock \bibinfo{journal}{Rev. Sci. Instrum.} \bibinfo{volume}{66}
  (\bibinfo{year}{1995}) \bibinfo{pages}{5486--5492}.
\bibitem[{Wilkins et~al.(1996)Wilkins, Gureyev, Gao, Pogany, and
  Stevenson}]{Wilkins1}
\bibinfo{author}{S.~W. Wilkins}, \bibinfo{author}{T.~E. Gureyev},
  \bibinfo{author}{D.~Gao}, \bibinfo{author}{A.~Pogany}, \bibinfo{author}{A.~W.
  Stevenson},
\newblock \bibinfo{title}{Phase-contrast imaging using polychromatic hard
  x-rays},
\newblock \bibinfo{journal}{Nature} \bibinfo{volume}{384}
  (\bibinfo{year}{1996}) \bibinfo{pages}{335--338}.
\bibitem[{Momose et~al.(2003)Momose, Kawamoto, Koyama, Hamaishi, Takai, and
  Suzuki}]{Momose3}
\bibinfo{author}{A.~Momose}, \bibinfo{author}{S.~Kawamoto},
  \bibinfo{author}{I.~Koyama}, \bibinfo{author}{Y.~Hamaishi},
  \bibinfo{author}{K.~Takai}, \bibinfo{author}{Y.~Suzuki},
\newblock \bibinfo{title}{Demonstration of x-ray talbot interferometry},
\newblock \bibinfo{journal}{Jpn. J. Appl. Phys.} \bibinfo{volume}{42}
  (\bibinfo{year}{2003}) \bibinfo{pages}{866--868}.
\bibitem[{Paganin et~al.(2002)Paganin, Mayo, Gureyev, Miller, and
  Wilkins}]{Paganin1}
\bibinfo{author}{D.~Paganin}, \bibinfo{author}{S.~Mayo}, \bibinfo{author}{T.~E.
  Gureyev}, \bibinfo{author}{P.~R. Miller}, \bibinfo{author}{S.~W. Wilkins},
\newblock \bibinfo{title}{Simultaneous phase and amplitude extraction from a
  single defocused image of a homogeneous object},
\newblock \bibinfo{journal}{J. Microsc.} \bibinfo{volume}{206}
  (\bibinfo{year}{2002}) \bibinfo{pages}{33--40}.
\bibitem[{Suzuki et~al.(2002)Suzuki, Yagi, and Uesugi}]{Suzuki1}
\bibinfo{author}{Y.~Suzuki}, \bibinfo{author}{N.~Yagi},
  \bibinfo{author}{K.~Uesugi},
\newblock \bibinfo{title}{X-ray refraction-enhanced imaging and a method for
  phase retrieval for a simple object},
\newblock \bibinfo{journal}{J. Synchrotron Radiation} \bibinfo{volume}{9}
  (\bibinfo{year}{2002}) \bibinfo{pages}{160--165}.
\bibitem[{Aoki and Kikuta(1974)}]{Aoki1}
\bibinfo{author}{S.~Aoki}, \bibinfo{author}{S.~Kikuta},
\newblock \bibinfo{title}{X-ray holographic microscopy},
\newblock \bibinfo{journal}{Jpn. J. Appl. Phys.} \bibinfo{volume}{13}
  (\bibinfo{year}{1974}) \bibinfo{pages}{1385}.
\bibitem[{Jacobsen et~al.(1990)Jacobsen, Howells, Kirz, and
  Rothman}]{Jacobsen1}
\bibinfo{author}{C.~Jacobsen}, \bibinfo{author}{M.~Howells},
  \bibinfo{author}{J.~Kirz}, \bibinfo{author}{S.~Rothman},
\newblock \bibinfo{title}{X-ray holographic microscopy using photoresists},
\newblock \bibinfo{journal}{J. Opt. Soc. Am.} \bibinfo{volume}{7}
  (\bibinfo{year}{1990}) \bibinfo{pages}{1847--1861}.
\bibitem[{MacNulty et~al.(1992)MacNulty, Kirz, Jacobsen, Anderson, Howells, and
  Kern}]{MacNulty1}
\bibinfo{author}{I.~MacNulty}, \bibinfo{author}{J.~Kirz},
  \bibinfo{author}{C.~Jacobsen}, \bibinfo{author}{E.~H. Anderson},
  \bibinfo{author}{M.~R. Howells}, \bibinfo{author}{D.~P. Kern},
\newblock \bibinfo{title}{High-resolution imaging by fourier transform x-ray
  holography},
\newblock \bibinfo{journal}{Science} \bibinfo{volume}{256}
  (\bibinfo{year}{1992}) \bibinfo{pages}{1009--1012}.
\bibitem[{Leitenberger and Snigirev(2001)}]{Leitenberger1}
\bibinfo{author}{W.~Leitenberger}, \bibinfo{author}{A.~Snigirev},
\newblock \bibinfo{title}{Microscopic imaging with high energy x-rays by
  fourier transform holography},
\newblock \bibinfo{journal}{J. Appl. Phys.} \bibinfo{volume}{90}
  (\bibinfo{year}{2001}) \bibinfo{pages}{538--544}.
\bibitem[{Suzuki et~al.(2010)Suzuki, Takeuchi, and Harada}]{Suzuki2}
\bibinfo{author}{Y.~Suzuki}, \bibinfo{author}{A.~Takeuchi},
  \bibinfo{author}{K.~Harada},
\newblock \bibinfo{title}{X-ray holographic microscopy by double-prism
  interferometer},
\newblock \bibinfo{journal}{Jpn. J. Appl. Phys.} \bibinfo{volume}{49}
  (\bibinfo{year}{2010}) \bibinfo{pages}{016601}.
\bibitem[{Suzuki and Takeuchi(2014)}]{Suzuki3}
\bibinfo{author}{Y.~Suzuki}, \bibinfo{author}{A.~Takeuchi},
\newblock \bibinfo{title}{X-ray refraction-enhanced imaging and a method for
  phase retrieval for a simple object},
\newblock \bibinfo{journal}{Jpn. J. Appl. Phys.} \bibinfo{volume}{53}
  (\bibinfo{year}{2014}) \bibinfo{pages}{122501}.
\bibitem[{Suzuki and Takeuchi(2008)}]{Suzuki4}
\bibinfo{author}{Y.~Suzuki}, \bibinfo{author}{A.~Takeuchi},
\newblock \bibinfo{title}{X-ray holographic microscopy using total-reflection
  mirror interferometer},
\newblock \bibinfo{journal}{Jpn. J. Appl. Phys.} \bibinfo{volume}{47}
  (\bibinfo{year}{2008}) \bibinfo{pages}{8595--8599}.
\bibitem[{Suzuki et~al.(2009)Suzuki, Takeuchi, and Uesugi}]{Suzuki5}
\bibinfo{author}{Y.~Suzuki}, \bibinfo{author}{A.~Takeuchi},
  \bibinfo{author}{K.~Uesugi},
\newblock \bibinfo{title}{Hard x-ray imaging holography and tomography},
\newblock \bibinfo{journal}{J. Phys.: Conf. Ser.} \bibinfo{volume}{186}
  (\bibinfo{year}{2009}) \bibinfo{pages}{012048}.
\bibitem[{Born and Wolf(1997)}]{Born1}
\bibinfo{author}{M.~Born}, \bibinfo{author}{E.~Wolf},
  \bibinfo{title}{Principles of Optics 6th edition},
  \bibinfo{publisher}{Cambridge University Press}, \bibinfo{year}{1997}.
\bibitem[{Watanabe and Aoki(2018)}]{Watanabe1}
\bibinfo{author}{N.~Watanabe}, \bibinfo{author}{S.~Aoki},
\newblock \bibinfo{title}{3d observation of quasicrystal alloy using x-ray
  differential phase-contrast microscope with a zone plate},
\newblock \bibinfo{journal}{Microsc. Microanal.} \bibinfo{volume}{24}
  (\bibinfo{year}{2018}) \bibinfo{pages}{166--167}.
\bibitem[{Arnison et~al.(2000)Arnison, Cogswell, Smith, Fekete, and
  Larkin}]{Hilbert1}
\bibinfo{author}{M.~R. Arnison}, \bibinfo{author}{C.~J. Cogswell},
  \bibinfo{author}{N.~I. Smith}, \bibinfo{author}{P.~W. Fekete},
  \bibinfo{author}{K.~G. Larkin},
\newblock \bibinfo{title}{Using the hilbert transform for 3d visualization of
  differential interference contrast microscope images},
\newblock \bibinfo{journal}{J. Microsc.} \bibinfo{volume}{199}
  (\bibinfo{year}{2000}) \bibinfo{pages}{79--84}.
\bibitem[{Wakabayashi et~al.(2022)Wakabayashi, Suzuki, Shibazaki, Sugiyama,
  Hirano, Nishimura, Hyodo, Igarashi, and Funamori}]{Wakabayashi1}
\bibinfo{author}{D.~Wakabayashi}, \bibinfo{author}{Y.~Suzuki},
  \bibinfo{author}{Y.~Shibazaki}, \bibinfo{author}{H.~Sugiyama},
  \bibinfo{author}{K.~Hirano}, \bibinfo{author}{R.~Nishimura},
  \bibinfo{author}{K.~Hyodo}, \bibinfo{author}{N.~Igarashi},
  \bibinfo{author}{N.~Funamori},
\newblock \bibinfo{title}{X-ray zooming microscopy with two fresnel zone
  plates},
\newblock \bibinfo{journal}{Re. Sci. Instrum.} \bibinfo{volume}{93}
  (\bibinfo{year}{2022}) \bibinfo{pages}{033701}.

\end{thebibliography}



\end{document}